# The wisdom of networks: A general adaptation and learning mechanism of complex systems

The network core triggers fast responses to known stimuli; innovations require the slow network periphery and are encoded by core-remodeling

*Peter Csermely*[*]

**I hypothesize that re-occurring prior experience of complex systems mobilizes a fast response, whose attractor is encoded by their strongly connected network core. In contrast, responses to novel stimuli are often slow and require the weakly connected network periphery. Upon repeated stimulus, peripheral network nodes remodel the network core that encodes the attractor of the new response. This "core-periphery learning" theory reviews and generalizes the heretofore fragmented knowledge on attractor formation by neural networks, periphery-driven innovation and a number of recent reports on the adaptation of protein, neuronal and social networks. The core-periphery learning theory may increase our understanding of signaling, memory formation, information encoding and decision-making processes. Moreover, the power of network periphery-related "wisdom of crowds" inventing creative, novel responses indicates that deliberative democracy is a slow yet efficient learning strategy developed as the success of a billion-year evolution.**

**Additional supporting information:**
Previously encoded ('fast', network core-related) and newly developing ('slow', network periphery-related) system responses are illustrated by three pairs of videos showing a neuronal network, as well as the social networks of network scientists and high school students.

Department of Medical Chemistry, Semmelweis University, Budapest, Hungary

**\*Corresponding author:**
Peter Csermely
E-mail: csermely.peter@med.semmelweis-univ.hu

# 1. Introduction

Complex systems often include a substantial number of components (referred to as nodes in their network representation). In case of random networks a large number of nodes would make the formation of system responses extremely slow and inefficient. To solve this problem, real-world networks often develop a network core that contains a few, densely connected nodes. The majority of nodes form the network periphery. Peripheral nodes are preferentially connected to the core, and are only sparsely connected to each other.[1–3] Network cores enable the development of fast and efficient responses (Box 1).[4]

The response-set of a complex system is encoded by its attractors. Attractors are defined states, in which the complex system converges.[5,6] In 1969, *Stuart Kauffman* described that random genetic control networks develop a surprisingly small number of attractors.[5] Recent studies indicated that rather few attractors represent the characteristic responses of a wide variety of complex systems including proteins,[7,8] cells,[9–12] neuronal[13–15] and social networks.[16,17]

Following the early work of *Little*, *Shaw*[18,19] and *Hopfield*[20] numerous studies of a rapidly growing field showed that learning processes of artificial intelligence networks lead to the development and consolidation of their attractors.[21] However, we know still surprisingly little about the mechanisms how biological networks encode new attractors as their responses to a novel challenge. From the 1980s a number of social science findings supported that creative innovations are often generated by the social/geographical network periphery, and core-periphery interactions play an essential role in their implementation.[22–27] Recently several lines of evidence related to protein structures, metabolic, signaling, neuronal, ecological and social networks have indicated that fast responses to known stimuli involve the network core, which drives the system to one of its attractors (see Section 2).[3,9,11,16]

From the above findings the following system-adaptation mechanism emerged. If the system experiences a novel challenge, the network core may fail to provide a coherent response; thus, the stimulus propagates to the periphery of the network. The novel response requires a substantial number of weakly connected peripheral nodes. If the novel challenge is repeated, the periphery remodels the core, and develops a new attractor. Core-remodeling may weaken or erase former attractors leading to 'forgetting'. However, the substantial knowledge supporting this mechanism remained fragmented (see Section 2).[1–27] In this paper I hypothesize that the above "core-periphery learning" response schema acts as an adaptation and learning mechanism of all complex systems, and will describe several potential ways of the development of novel system attractors.

# 2. Examples of system response duality to well-known *versus* novel environmental changes

Prior to describing the core-periphery learning hypothesis of this paper in detail, I first list a set of salient examples that indicate the duality of previously encoded *versus* newly acquired system responses.

## 2.1. Protein structures

Proteins may be described as residue interaction networks, where nodes are amino acids and network edges connect adjacent amino acids in the 3D protein structure. The core of protein structure networks is enriched in hydrophobic amino acids.[3] Importantly, the network periphery often contains intrinsically disordered protein segments, which occur in 85% of human regulatory proteins.[28,29]

The protein core provides an exceptionally fast energy-transfer, which is localized to a few key amino acids as demonstrated by experimental findings, molecular dynamic simulations and analysis of evolutionary conserved sequences.[30–37] The protein core is tightly packed, where conflicting constraints and forces lead to the 'frustration' of several amino acid residues.[38] Analysis of frustration and rigidity changes may help the identification of core segments mediating the fast transmission of well-known environmental changes in protein structures.[39–41] Allosteric activation often induces inter-microdomain coupling, which expands the protein core and makes overall energy transfer faster and more efficient.[7,37,42] Thus allostery introduces an 'on-line' sensor of 'well-known' environmental changes.[8]

The example of proteins is somewhat special, since protein responses to well-known environmental changes were developed by evolutionary selection processes long time ago. However, peripheral, intrinsically disordered segments enable the fast adaptation of proteins to environmental changes. Disordered segments often fold upon binding to the upstream signaling partner, allowing 'conformational signaling' as exemplified by the nuclear hormone receptor family and the cAMP response element-binding protein (CREB) among many others.[29] Folding of intrinsically disordered segments often occurs in the range of seconds,[43] which is much slower than either the picosec time-scale[30,32,37] of protein core-mediated signaling, or the nsec to msec timescale of conformational changes.[29]

Conformational memory gives yet another example of environment-induced protein dynamics. *E. coli* lactose permease displays a lipid-induced conformational change, which remained detectable well after the removal of the lipid.[44] Similarly, human glucokinase,[45] or the ATP binding cassette transporter, BtuC$_2$D$_2$,[46] retained a hyper-activated state long after the completion of their catalytic cycle. These kinetic forms of allostery may represent a general feature of many proteins that was introduced as allokairy.[47] Self-templating conformations of prions and an increasing number of other, intrinsically disordered, evolutionarily conserved proteins induce heritable traits.[48] In addition, prion-like proteins were first hypothesized,[49] and were subsequently shown to be involved in the memory storage of neurons.[50,51]

In summary, protein cores show a fast and evolutionarily optimized signal transmission concentrated to a few amino acids. On the contrary, peripheral, intrinsically disordered protein segments provide a wide range of slower responses including their upstream-signaling partner-induced folding and conformational memory.

## 2.2. Metabolic and signaling networks

Metabolic networks are the network representation of cellular metabolism, in which enzyme proteins are the edges and their substrates are the nodes. The minimal gene set-related, evolutionarily conserved core and the environment-dependent periphery of metabolic networks[9,52,53] exhibit a very similar duality to that observed in protein structures. This network structure stabilizes metabolic attractors, which are correctly recovered after a matching input.[9]

Cellular signaling network states also converge at attractors, which are re-configured after environmental changes.[5,10–12,54] Importantly, cell differentiation may be modeled by the activation of core gene expression processes, which drive the shift between the major attractors of progenitor and differentiated cells. In addition, a substantial number of transient, peripheral gene expression processes capture the pathways and responses specific to the actual input.[54]

Both metabolic and signaling networks can be reconfigured by signal-directed folding of intrinsically disordered proteins,[29] as well as by chromatin-related, epigenetic learning mechanisms providing an enhanced response after a repeated stimulus.[55–57] It requires further studies, whether these direct and indirect 'learning' processes preferentially modify network peripheries, as expected. Our current knowledge on previously encoded and newly formed responses of protein structure, metabolic and signaling networks is summarized in Table 1.

## 2.3. Neuronal networks: connectomes

This set of examples will describe the responses of neuronal networks to well-known and novel stimuli. Nodes of neuronal networks (connectomes) are neurons. Inter-neuronal edges are primarily provided by synapses.

In pyramidal, place sensing neurons of the rat hippocampus several highly active, fast-firing neurons are surrounded by neurons that are less active and slower. Fast neuronal matching is often imprecise, which increasingly occurs, when the rat experiences new locations. In this latter case, spatial map refinement by the giant, weakly connected network of most neurons becomes particularly important.[58] Importantly, a smaller subset of slow-firing, plastic cells gains high place-specificity during exploration and exhibits increased bursting and co-activation during post-experience sleep.[59] Thus, plastic neurons involved in the precise encoding of novel stimuli become similar to the rigid neurons that encode previous information. Similarly, an enriched environment sensitizes mouse dentate gyrus granule cells, which enables their fast conversion to highly excitable and tightly connected network cores that encode new information.[60] Similar core-periphery task differences could be observed in several other neuronal structures as shown below.

➢ A substantial part of rat motor cortex may belongs to the giant, weakly connected network periphery involved in learning novel motor tasks. In contrast, motor cortex lesions had no discernible effect on previously acquired motor skills.[61]
➢ Task-relevant visual areas exhibited a higher-than-average topological proximity to the network core in a relatively low resolution functional magnetic resonance imaging analysis of human brains. The core-association of task-relevant visual areas was further increased during correct visual task solutions.[62]
➢ The core of 34 mushroom body output neurons (a locus where the 2000 Kenyon cells of the *Drosophila* olfactory system converge) encoded various odors in a highly correlated manner. In contrast, the peripheral layers of sensory input remained highly decorrelated.[63]
➢ As a final, and perhaps most convincing example, sensitization of the escape swim of the marine mollusk, *Tritonia diomedea* recruited peripheral neurons to the reliably bursting neurons of the network core, which increased the vigor of the elicited swim response.[64]

Memory retrieval provides an excellent example for the mobilization of a previously encoded response by a "well-known," re-occurring situation. A recent study has indicated that repetitive activation of the visual cortex of awake mice built newly developed, stable neuronal core structures that were mobilized together after being imprinted.[65] Similarly, a core of several highly correlated, hub-like neurons was formed during the mouse fear memory learning process. In fear memory retrieval, core neurons tended to lead their correlated neuron pairs in the network periphery, which resulted in network-wide synchronous events. This finding indicated that the neuronal network core acts as an 'opinion leader' initiating responses to known stimuli.[66]

Going beyond structural findings on neuronal networks *Daniel Kahneman* described[67] that fast thinking determines our own actions if we encounter a familiar situation. However, if an event occurs that violates the model of the world encoded by our fast thinking neuronal system, the slow, contemplative, 'deep' thinking system becomes activated, which enables more detailed and more specific processing. This division of labor is highly efficient in minimizing effort and maximizing performance. In the initial phase of the response fast and slow thinking often complement each other. The fast opinion is implemented if immediate action is needed, while preparing a slow, deep thinking solution as the refinement.[67] This description summarizes a massive number of behavioral studies. Examination of the role of whole-brain connectome core and periphery in fast and slow thinking processes will be an exciting task of future studies.

In conclusion, a substantial number of corroborating findings[58–67] suggests that memory is encoded by the extension of a fast-responding network core of correlated neurons that recruit and activate neurons from the network periphery. In memory retrieval processes the previously formed core reacts first and elicits a general response of the periphery. Notably, the relatively small number of peripheral neurons involved during memory formation[59,60,64,66] may reflect current experimental limitations because it is a daunting task to obtain individual records of a substantial number of weakly connected/correlated neurons, whose sporadic activation by a novel situation may precede the recruitment of a limited number of peripheral neurons to the network core.

## 2.4. Ecosystems

The general and specific resilience of ecosystems against previously experienced *versus* novel changes[68,69] indicates a highly similar response duality to the changes described in the previous sections. As an interesting example during the continuous reconfiguration process of an arctic seasonal pollinator community, a high ecosystem structural stability was reached *via* the incorporation of the continuous flow of newcomer species to the network core.[70]

Importantly, recent work on ecosystem evolution showed that a memory of the phenotypes that have been selected in the past can facilitate faster adaptation, whenever these phenotypes are selected again in the future. Moreover, ecosystem memory can also facilitate faster adaptation to new challenges by recombination of previously learned solutions.[71]

## 2.5. Social networks

Finally, I describe the decision-making mechanisms of social networks. Well-known individuals often know each other forming a tightly connected network core. The core gives fast responses to previously occurring situations but often forms echo-chambers resulting in a significant confirmation bias. In contrast, the network periphery provides a wider range of solutions than the core and can overturn suboptimal choices of the social 'elite' (Box 2).[3,72–77] This core-periphery behavior underlies the importance of 1.) the expansion of the definition of expertise; 2.) creation of a culture, that is truly receptive to new ideas, and 3.) the empowerment of opportunity scouts.[78] The weakly connected majority of the network periphery has a key role in the collective opinion of *James Surowiecki*'s proverbial *"Wisdom of crowds"*.[76] Thus, the development of the optimal response requires the contemplative thinking of the entire community that uses the inclusive, self-governing and citizen-powered processes of deliberative, slow democracy.[79,80]

In conclusion, the wide range of examples listed in this paper (Table 1) strongly suggests that the development of novel, optimal responses requires the contribution of the whole community. Importantly, this can not be perceived only as the vote of the majority, especially, if votes were casted without a previous extensive discussion of the subject, but were based on simplified slogans or 'identities'. The process of deliberative democracy raises the 'crowd' to the level of experts, which is in agreement with *Plato*'s ideas on a well-functioning democracy.[81]

# 3. The core-periphery learning hypothesis: Known stimuli trigger fast responses of the network core, whereas the development of novel attractors requires the network periphery

From the wide range of examples described in Section 2 (for a summary: see Table 1), a dual response-pattern of complex systems emerges. Generalizing this pattern, here I describe the development of novel attractors as the hypothesis of the current paper.

As a starting step, the stimulus reaches the network core in a fast process as shown by the red nodes on Fig. 1 and by the illustrative videos of the additional supporting information.[82–85] This is expected, because core nodes typically have a substantially higher number of neighbors than other nodes, and are connected with edges that have a large weight (solid lines of Fig. 1; Box 1). After this starting step one of the following three scenarios may happen.

## 3.1. Scenario 1: Shift to an attractor encoded by previously encountered situations

As key findings supporting my hypothesis several recent publications proved that the node sets (called as stable motifs or feedback vertex sets)[86–90] that determine system attractors are part of the strongly connected components of directed networks (where every node is reachable from every other node). The strongly connected component is, in fact, the mathematically defined core of bow-tie networks (Box 1). These studies showed that 1.) core nodes play a major role in encoding system attractors; 2.) different attractors may be encoded by overlapping subsets of core nodes and 3.) not all core nodes are participating in attractor-encoding.[14,86–90] Since the initial form of my hypothesis was published as a preprint,[91] several other supporting findings have appeared[59,60,64,65,70,72,90] that support and extend the original concept. Taking the above studies and the list of examples before together, I generalize the following scenario.

If the incoming stimulus had been experienced by the complex system several times before, a set of core nodes have formed a group, which drove the system to an attractor giving an adequate response to the stimulus. If now the same stimulus is repeated again, the system is driven to this attractor (Fig. 1A). This provides a fast, reliable and robust response (for time scales, see Table 1).

Peripheral nodes (forming the "in" and "out" components of bow-tie networks) may refine the form and size of the attractor basins but may not influence the number of attractors. (Please note, that here the attractor may also be a set of fixed points, a limit cycle, or a limit torus.)

## 3.2. Scenario 2: System response to novel situations

If the stimulus originates from a novel, unexpected situation (Fig. 1B) it may be incompatible with the existing attractors set by the core. Thus, the stimulus may provoke conflicting core responses inducing the system to fluctuate between its original attractors. Here the stimulus propagates to the majority of weakly linked peripheral nodes, which stabilize the system. Besides system stabilization peripheral node involvement enables the emergence of slow (yet creative) responses to the novel stimulus as the 'collective decision' of (practically) the entire network (for time scales, see Table 1). This process may modify the position, size, saddles or depth of the complex system's attractor basins. Note, that the emergent periphery-response is slow not only because the re-organization of the periphery is requiring a large number of rather slow, stochastic steps (as detailed in Section 4), but also because stimulus-driven periphery reorganization must often be attempted hundreds (if not thousands) of times before finding a new, adequate response.

### 3.3. Scenario 3: Encoding a novel system attractor

In case the novel stimulus is repeated (many times), the peripheral network nodes, which were involved in "Scenario 2", may gradually reconfigure the network core adding nodes to it, or exchanging its nodes (Fig. 1C). This process encodes the newly acquired response as a novel attractor of the system. Core-reconfiguration may weaken or erase some of the earlier system attractors and thus may also serve as a 'forgetting' mechanism.

    This core-periphery learning hypothesis is novel, since it connects previously fragmented knowledge on the capability of model networks to develop system attractors in learning processes[18–21] on network periphery-generated creative innovations[22–27] and on the differential role of core and periphery nodes in attractor formation in a large variety of real-world networks.[3,9,11,16,29–90] Moreover, the hypothesis expands these lines of evidence to a general adaptation and learning mechanism of complex systems.

    The rigorous proof of several details of the above three scenarios requires additional studies. As an extension of the core-periphery learning hypothesis I will detail the potential mechanisms of Scenarios 2 and 3 (development and encoding of novel system responses) in the next section. Limitations, several possible further proofs and potential applications of the hypothesis will be described in subsequent sections of this paper.

## 4. Development and encoding of novel complex system responses as potential ways of adaptation and learning

What is the mechanism of the formation and consolidation of new system-responses? How do 'creativity', 'deep thinking', 'contemplation' and 'deliberation' emerge in biological systems? How are optimal responses encoded to the complex system's network so they may be efficiently retrieved later? As the extension of the core-periphery learning hypothesis described in the previous section, here I seek answers to these questions and describe potentially general, system-level adaptation and learning mechanisms.

    One of the most persuasive learning mechanisms involves the increase of synaptic strength between neurons.[50] Importantly, this *Hebbian* learning rule is related to the network core reconfiguration of "Scenario 3": if an edge-weight increases as the system encodes a novel response, the nodes that belong to this edge may become a part of the reconfigured network core connected by high-weight edges (Box 1). Notably, neuronal learning may involve several mechanisms other than the increase of synaptic strength, such as changes in bursting behavior, excitability or the structure of perineuronal nets.[50,65,92,93] In addition, stimulus-mediated edge weight increases of signaling or metabolic networks have not been fully established yet, but may involve signaling-induced folding of intrinsically disordered protein regions[29] or epigenetic memory-related learning mechanisms providing an enhanced response after a repeated stimulus.[55–57] Edge-weight remodeling re-channels the information flow in a network. Re-channeling appears to be applicable as a general learning mechanism and may involve several cases as subsequently described.

## 4.1. Re-channeling of information flow by connecting distant network regions using creative nodes

A drastic re-channeling of information flow may be achieved if re-channeling connects formerly 'quasi-distinct' regions of the complex system's network. These regions must reside in the sparsely connected network periphery (and *not* in the densely connected core). Highly dynamic, weakly linked nodes that connect various distant network regions have previously been termed "creative nodes".[94] Creative nodes bridge the "structural holes" of *Ronald Burt*.[95] Various forms of creative nodes in different networks are listed in Table 1. The increase of creative node edge-weights may be a particularly suitable method to remodel system attractors. *Henri Poincaré* defined creativity as connecting distant regions of human knowledge as follows: "*to create consists in not making useless combinations.... Among chosen combinations, the most fertile will often be those formed of elements drawn from domains which are far apart*".[96] In agreement with this statement, an analysis of 17.9 million scientific papers showed that highest-impact papers feature an intrusion of unusual combinations.[97] In recent social experiments and simulations, the accumulation of high-complexity innovations required both the separation and occasional connection of distant groups resulting in creative combinations.[77] In an extensive study of Facebook comments, significantly greater attention was triggered by messages that combined themes seldom discussed together. These "cultural bridges" often induced new conversational themes that acted as "cultural trellises".[98] Similarly, Wikipedia users prefer links pointing towards the periphery of the Wikipedia network[99] indicating a search for novelty not in the redundant core, but in the non-redundant periphery. These findings regarding creativity are in agreement with the re-channeling of information flow by connection of the distant network regions described here.

## 4.2. Re-channeling of information flow by changing edge directions

Information flow can be efficiently re-channeled by changing the direction of even a single edge. This little change may drive the system from a hierarchical control by a limited number of nodes to a community-control by the majority of nodes.[100] This duality is closely related to the core-periphery duality of the previously described response pattern. Edge-direction change may be triggered by the decrease in the rigidity of the more rigid node (Box 3).[101] Importantly, change in the direction of an edge may introduce loops in directed networks. This may dramatically increase their plasticity, and may destabilize/reconfigure their former attractors.[102]

## 4.3. Re-channeling of information flow by core remodeling as a way to encode new attractors

Finally, I will describe three mechanisms that remodel the network core and encode novel attractors. Importantly, core remodeling may also erase part of the previously encoded system responses (attractors), which may thus also be a mechanism of forgetting and consequent system reset.

*4.3.1. Mechanism of network core remodeling 1: Core-conflict mediation by 'creative' or 'innovator' nodes*

In case of a novel stimulus, core nodes often trigger different responses. Contradicting responses induce a fluctuation between previously encoded attractors. Here, 'mediation' of contradicting core responses becomes useful. This 'mediation' is often provided by core-adjacent, non-hub nodes. Species of the schooling fish, *Notemigonus crysoleucas* with relatively few, yet strongly connected, neighbors were both most influential and most susceptible to social influence.[103] The high influence of inter-hub nodes was also demonstrated in large-scale social networks of Twitter or mobile phones.[104] Notably, mediating nodes often have weak links that resemble to those of the "creative nodes" previously described.[94] Thus, core-conflict mediation may contribute to the stabilizing "strength of weak ties" initially described for social networks by *Mark Granovetter*[105] and subsequently generalized to many complex systems.[106]

In the concept of innovation diffusion mediator nodes correspond to "innovators" (bridges, brokers). Here, the opinion leader, socially integrated "early adopters" predominantly belong to the highly active network core.[23,27] Creative innovations are often generated by the network periphery.[22,24,25] The network periphery is a preferred position of innovators because they may have contacts here with other social communities and may be free from the social pressure of core-enforcing conformity. Importantly, external advisors, consultants and change agents typically occupy the position of the hub-connecting, core-interacting, low-degree nodes, which partially explains their highly influential status. Core-members are typically afraid of changing the status quo, which may jeopardize their prestigious position. Thus, traditionally behaving core members seldom become innovators.[23,27] Importantly, core-periphery interactions play an essential role in the spread and implementation of innovations.[25,26]

*4.3.2. Mechanism of network core remodeling 2: Core-reconfiguration by addition and/or exchange of core nodes*

Repeated stimuli may transform core-mediation to core-reconfiguration, in which core-associated, mediating nodes become part of the core that encodes the novel response.[59,60,64,65] The core may also lose some of its previous nodes during reconfiguration, as shown in neuronal and social networks.[3,64] These examples indicate that core-reconfiguration may induce the weakening/erasure of former system responses during the encoding of a novel attractor. Thus, core-reconfiguration also represents a forgetting mechanism.

*4.3.3. Mechanism of network core remodeling 3: Core-melting*

A mismatched stimulus may also 'melt' (and thus erase) part of the core by a decrease of its edge weights and rigidity.[39,41] Increased plasticity helps to generate novel attractors and/or makes existing attractors accessible. If the mismatched stimulus is repeated it may encode a novel set of constraints to the network structure establishing a new segment of the network core. This core-extension makes the network more rigid again. These plasticity/rigidity cycles characterize a wide range of adaptive processes.[40,41] Similarly to core-remodeling discussed in the previous paragraph core-extension (resembling to the 'election of new leaders') may also enrich the system with a newly encoded attractor.

Importantly, 'core-melting' may represent a key mechanism of forgetting. The slow relaxation of high-affinity, high-turnover protein conformations after the dissociation of the substrate[45–47] is a molecular-level model of forgetting. A similar decrease in the network rigidity may induce forgetting at the structure of the actin cytoskeleton in neuronal synapses,[107] as well as at the perineuronal net.[92]

The above three mechanisms (core-mediation, core-reconfiguration and core-melting) indicated that minor changes in the network core may lead to gross changes in network behavior. This is rather plausible because the core often determines the major system attractors. However, the network periphery may remodel attractor basins, make attractors accessible or inaccessible, merge two attractors or divide a large attractor basin to several smaller ones. In case of repeated stimuli, these periphery effects may reconfigure the core and encode novel attractors.

These examples showed that learning mechanisms can be extended from neuronal networks to other complex systems, such as protein structures (especially those of intrinsically disordered proteins), metabolic networks, ecosystems and social communities. This assumption is in agreement with the recent concept of *Watson & Szathmáry*,[71] who described the evolutionary adaptation of ecosystems as a learning process.

## 5. Limitations, potential proofs and applications of the core-periphery learning hypothesis

This section will list the limitations of the core-periphery learning hypothesis giving a few exceptions of the general adaptation mechanism described in this paper. However, as evidenced by the salient examples listed before, these exceptions, by far, do not represent the majority of cases.

- Importantly, the speed of response itself may not always discriminate core or periphery (formerly encoded or newly learned) responses. Responses may become slow either because the stimulus removed the system far from its corresponding attractor, or because the formation of its corresponding attractor requires a long time. However, the latter process usually requires even longer time that the former, since periphery reconfiguration is stochastic and requires many repeats (see Table 1).
- In some emergency scenarios, complex systems may adapt with an initial, approximate, mismatched, yet fast response of their core, which subsequently becomes refined by the slower contributions of the periphery.[67] Important scenarios of this response are, when the system remains fluctuating in a bistable switch (such as the optical illusion of the Necker cube) in a limit cycle or a limit torus.
- Some 'simple' systems, such as protein structures lacking intrinsically disordered regions, may not be able to 'learn' novel responses. However, they reflect the constraints of previous evolutionary selection steps.
- Highly specified, engineered networks may often lack adaptive responses.
- Core/periphery fluctuations may occur in some of those neuronal and social networks that display a high plasticity.[3,41,64]
- The "wisdom of crowds"[76] may be converted to the "madness of crowds",[108] thereby leading to widespread popular delusions. Creative nodes[94] may prevent these catastrophes. Various forms of the "madness of crowds" phenomenon and creative nodes are listed in Table 1.

Several limitations of the core-periphery learning hypothesis are, actually, its extensions.
- In some cases, the core may have an excessive number of constraints. This extreme core rigidity severely limits the core's portfolio of fast responses. These super-rigid cores may reject most stimuli and may appear 'purposefully slow', similar to bureaucracies.
- The core may have multiple segments (Box 1). As an example of the functional utility of a dissociated network core forming network modules, the interconnection of brain modules has been identified as a key process of human cognitive functions.[109]
- Learning may proceed in repeated cycles with varying contributions of distant network segment reconnections, edge reversals and core remodeling.[39,41]

Potential proofs that may support (or may rule out) the core-periphery learning hypothesis are listed in Box 4. The dual adaptive mechanism described in this paper contributes to our understanding of signaling, learning and decision-making processes. Innovations resembling to core-periphery remodeling[110–112] may overcome the overfitting, slow convergence, "fooling effect" and catastrophic forgetting of several current artificial intelligence techniques. Recent Internet innovations recognize the importance of Internet periphery to adapt to variable challenges[113] and involve the design of an adaptable Internet core.[114] Finally, network-based drug design has recently emerged as a novel paradigm of drug development.[40] While the "central hit drug design strategy" targets the network core,[40] the "network influence drug design strategy" targets peripheral nodes,[40] preferably hitting those nodes that are similar to the highly susceptible, highly influential, hub-connecting, core-adjacent nodes described in this paper.

## 6. Conclusions

In conclusion, a wide range of evidence indicates that upon an environmental stimulus complex systems mobilize a fast, pre-set response of their well-connected network core shifting the system to one of its attractors. If this fails, the stimulus propagates to the weakly connected network periphery, and a slow, integrative response of the entire system develops. In case of repeated stimulus this integrative response may remodel the network core and encode a novel attractor. Thus, a wide range of natural systems mobilize their network periphery and initiate a process similar to 'deliberative, deep thinking' when creating novel responses. Further studies on core- and periphery-driven responses will give more insight into adaptation and learning mechanisms, as well as construct more efficient future technologies.

The generality of the "wisdom of crowds" described by the core-periphery learning hypothesis here indicates that deliberative democracy is an efficient learning strategy optimized by complex systems as response to unexpected situations in billion-years of evolution. The 21$^{st}$ century is full of novel situations that have not been previously experienced by mankind. This paper warns that we must put substantially more effort into mobilizing the hidden wisdom of our human communities and the deep thinking of creative, talented minds to survive these challenges.

**Acknowledgments**
This paper is dedicated to the memory of George Klein (working at the Karolinska Institute, Stockholm, Sweden for almost 70 consecutive years), who gave important encouragement and advice for this paper, and whose inspiration and wisdom is missed and will be kept by many. The author is thankful Balázs Baksa (http://albafilm.hu) and Máté Szalay-Bekő, who provided the video illustrations. The criticism, suggestions and advice of members of the LINK-Group (http://linkgroup.hu), particularly István Kovács, Kristóf Szalay and Zsuzsa Szvetelszky, as well as the late George Klein (Karolinska Institute, Stockholm, Sweden), Réka Albert (Pennsylvania State University, State College, PA USA), György Buzsáki (New York University Langone Center, New York, NY USA), Tamás Freund, Balázs Hangya, Zoltán Nusser (Institute of Experimental Medicine, Hungarian Academy of Sciences, Budapest, Hungary), Robert May (University of Oxford, Oxford UK), Balázs Papp (Szeged Biological Research Centre, Hungary), Péter Tompa (Vrije Universiteit, Brussels, Belgium and Hungarian Academy of Sciences, Budapest, Hungary), Gábor Tusnády (Alfréd Rényi Institute of Mathematics, Hungarian Academy of Sciences, Budapest, Hungary) and Peter Wolynes (Rice University, Houston TX USA), are gratefully acknowledged. This work was supported by the Hungarian National Research, Development and Innovation Office (K115378).

The author has declared no conflicts of interest.

**Box 1**
**Definition and properties of the network core and periphery.** The network core refers to a central and densely connected set of a few network nodes, where connection density is often increased further by large edge weights. In contrast, nodes of the network periphery are non-central, sparsely connected, and attach preferentially to the core.[1–3] Rich-clubs represent interconnected hubs, which may form a part of the network core.[2] Importantly, the strongly connected component of directed networks (where every node is reachable from every other node) is the mathematically defined core of their bow-tie structures. The mathematically precise definitions of other core concepts are listed in Ref.[3] The core of modular networks is composed of multiple, densely connected regions.[3] Both single and multiple network cores have been shown to stabilize complex systems by the early work of *Robert May*.[4] The network core provides a plausible structure to store previously encoded system responses because it is central and easily approachable, yet simultaneously shielded from the environment by the network periphery. Core-shielding is evident in protein structures where the network core is the physical core of the protein that contains hydrophobic amino acids and is shielded from the surrounding water by the peripheral amino acids. Shielding of signaling and neuronal networks protects them from over-excitation by prolonged stimuli. Moreover, super-influential members of the social elite tend to shield themselves from direct public influence by imposing tight control of their appearances in publicly open situations. Besides core-shielding, the preservation of system responses is helped by the evolutionary conservation of network cores, since the dense connections of the core impose a set of system-constraints. Notably, this set of system-constraints is exactly the information the core has preserved when the system's optimized attractor repertoire was set by previously encountered situations.

**Box 2**
**Roles of the 'elite' and the 'wisdom of crowds' in decision-making processes.**
The core-forming elite may trigger and lead an efficient and fast response of the entire community if the challenge was previously experienced and/or trivial. In contrast, after a novel, unexpected challenge, the development of an adequate response requires the variability of the flexible majority of the network periphery, i.e., the "wisdom of crowds".[3,72–77] Several examples demonstrate this duality. Even in chacma baboon groups routine group movements are driven by the network core.[72] The small Twitter network core produces the majority of tweets, which are characterized by mobilizing, polarized political views. In contrast, tweets of the periphery reflect 'contemplative', politically more moderate views.[73] In the widely used voter model the 'wrong' (i.e., less preferred) opinion of the 'top-leader' of a directed, perfectly hierarchical tree network was overturned by the 'right' (more preferred) public opinion – as distant network nodes became connected.[74] The increase of randomness also reduced the appearance of 'extremism' and increased 'deliberative thinking' in a different model of collective opinion formation.[75] Furthermore, the effects of highly confident, core-type individuals and the majority of laypeople have been shown to act as the two major attractors of a group's opinion in controlled experiments.[16] The "wisdom of crowds"[76] has been further demonstrated by examples of 'human computation', in which an extensive number of participants worked independently with rules encouraging the generation of new insights.[77]

**Box 3**
**A potential mechanism that changes the direction of a network edge.** As described in the main text, changing the direction of even a single network edge may re-channel the network information flow and change the system behavior.[100] 1.) How is the edge-direction formed in nature? The "induced-fit" mechanism of protein interactions, in which the more rigid partner influences the less rigid partner,[8] provides a rather plausible rationale of edge-direction formation. Note that this edge-direction definition remains valid if we perceive rigidity as functional rigidity,[39] which indicates that the less rigid node (which is, in most cases, a complex system itself, such as a single neuron in neuronal networks) has a substantially higher number of attractors than the more rigid node. Thus, the less rigid node, having more attractors, has much greater chances of accommodating itself to the actual status of the more rigid node than vice versa. In case the more rigid node 'melts', i.e., decreases its rigidity below that of the less rigid node, the direction of the edge becomes reversed. 2.) How can the more rigid node be 'melted'? This may be achieved by a concentration of energy on the more rigid node, which may re-arrange its inner structure in a way that it becomes more random, noisier, or more plastic. This assumption is plausible because increased resources lead to increased randomness of network structures,[101] which causes their increased plasticity.[39,41] 3.) How does the energy received by the network become concentrated on the more rigid node? Observations of protein structures have indicated that the energy tends to accumulate at the most rigid segments of the protein.[36] This is plausible, since rigid segments preserve and transmit signals better, whereas plastic segments dissipate them better.[39–41]

**Box 4**
**Potential proofs to support (or to rule out) the core-periphery learning hypothesis.** Here I list some experiments that may give further support for the core-periphery learning hypothesis described in this paper (or may highlight its limitations beyond those listed in Section 5 of the main text).
- Detailed kinetic studies of signaling-induced folding of intrinsically disordered protein segments should reveal their conformational memory remaining transiently folded after dissociation from their partner, and thus sensitizing the cellular response to a repeated stimulus.
- Intrinsically disordered proteins should be enriched in the periphery of signaling networks (as opposed to their core).
- Signaling responses to well-known and novel stimuli have not been readily discriminated yet. This is partially due to an experimental bias, since we usually expose the cell to a single stimulus. Future experiments that add 'previously-experienced' or novel 'cocktails' of hormones, cytokines, etc. and measure system-wide signaling responses may allow the discrimination of core- and periphery-centered signaling.
- I expect a lot more studies revealing "intergenerational memory" of cellular signaling and metabolic responses than the pioneering paper of *Doncic* et al.[117] and the initial findings on epigenetic memory.[57]
- The refinement of neuronal techniques may provide additional evidence for the key role of weakly connected, peripheral neurons[59,60,64,66] in the development of novel neuronal responses.
- Whole-brain connectome studies should reveal that fast and slow thinking processes[67] are related to the connectome core and periphery, respectively.
- The differences between network core- and periphery-induced ecosystem reconfigurations are largely unexplored. Future studies should reveal more connections between ecosystem network cores and ecosystem memory.

**Table 1. Comparison of previously encoded and newly formed responses of various complex systems**

| System | System's core / periphery | System's attractor | Incoming stimulus / its activation | Previously encoded response | Newly formed ("wisdom of crowds" type) response | "Madness of crowds" response / creative nodes* | Time scale of previously encoded / new responses** |
|---|---|---|---|---|---|---|---|
| **Protein structure networks** | tightly packed / intrinsically disordered segments | conformational state | ligand or protein (RNA/DNA) binding / allostery, intrinsically disordered segment folding/unfolding[29,94] | evolutionary optimized protein structure, conformational memory | induced folding/unfolding of intrinsically disordered segments[29] | (partial) misfolding, aggregation / discrete breathers, hot spots[29,36,94,115] | psec,[30,32,37] nsec to msec (in case of conformational change)[29] / msec to sec[43] |
| **Metabolic and signaling networks** | housekeeping, maintenance / environment-dependent processes | cellular phenotype | molecular signal / protein activation, inhibition or translocation | responses adequate to the cellular phenotype and its epigenetic memory | responses characteristic to a different cell(ular pheno)types | diseases (e.g. cancer) / date hubs, chaperones[94] critical signaling nodes[116] | sec to min / hours[55,56,117] |
| **Neuronal networks** | persistently co-activated / sporadically associated neurons | thoughts, individual behavior | sensory or cognitive input / neuronal firing | memory, imprint, skills, engram + their retrieval | exploration, creativity, discovery, learning | hallucinations, illusions / dentate gyrus neural stem cells[60] | msec range / msec to sec[50,93] |
| **Ecosystems** | generalists / specialists[70] | ecosystem dynamic regime: collective behavior[70,71] | population change (e.g. invasion, predation, stress) / ecosystem reconfiguration | ecosystem memory and general resilience[68,69,71] | core-reconfiguration, emergence of specific resilience[68–70] | runaway, microbiome revolt[118] / omnivores, top predators | N.A.*** |
| **Social networks** | elite / social periphery, minorities | group- and society-level response | new information / information spread | social memory: customs, rules, norms, culture | creative, novel solutions at the community level | spread of delusions / avant-garde minority, connector, consultant, market guru, prophet | sec to min / hours to years |

*The "madness of crowds" phenomenon occurs, when the system (like a cell) get stuck in its own optimum, which does not match with the optimum of the meta-system one level higher in the hierarchy (like the organism made by the cells). Creative nodes are highly dynamic network nodes having a very unpredictable behavior, which help the system to find its optimum matching with its environment.[94] Due to space limitations the table gives only a few examples of the possible many for both.

**Note that these time scales are rough estimates, and may vary. The previously encoded stimulus may drive the system far from its response-encoding attractor and thus the encoded response may develop slower than these estimates. Consolidation of the newly encoded stimulus may need several network reorganization attempts and may also develop much slower, than these estimates. Importantly, in most cases currently we have not enough data to give an estimate for the time scale of the network re-arrangements inducing newly formed responses. Therefore the time scale of the network rearrangements was not included to this Table.

***N.A. = ecosystem time-scales depend on the life-cycle of participating specii and the size of the ecosystem so much, which precludes a general estimate here.

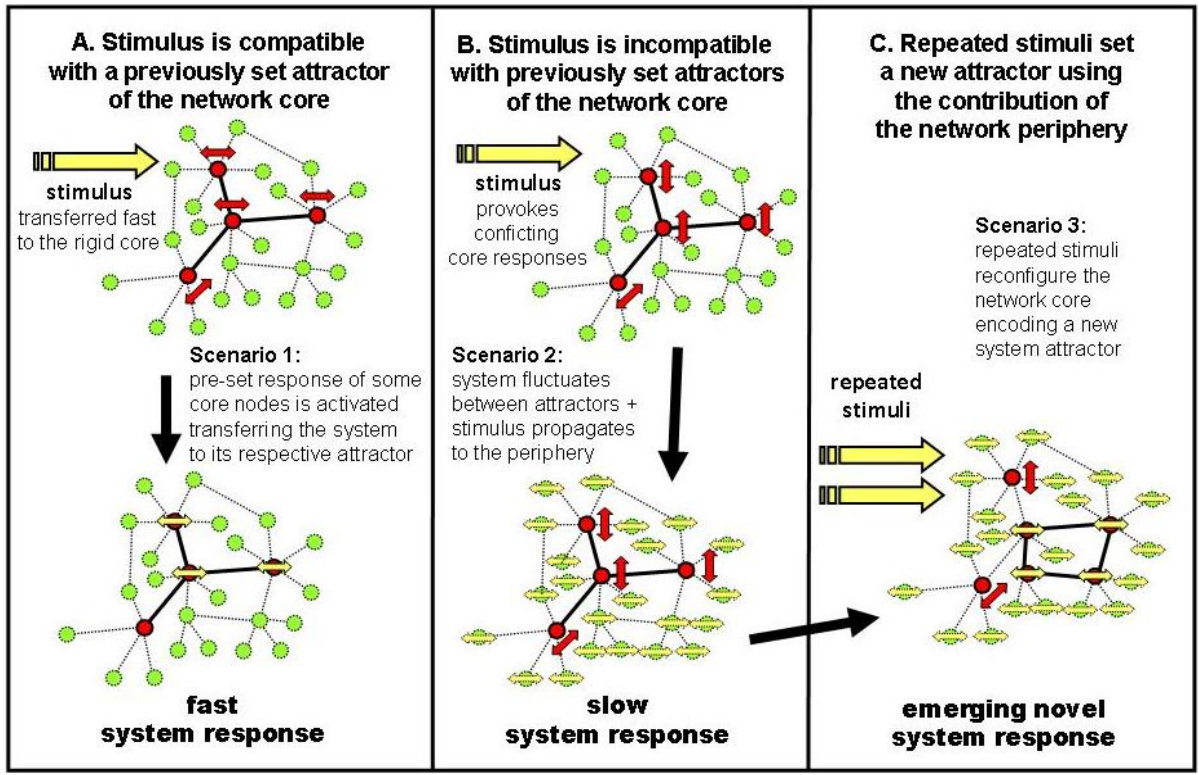

**Figure 1.** Description of the core-periphery learning mechanism of complex systems. The stimulus is rapidly concentrated at the rigid core of the network (*red nodes*) as a result of the core's central position and large weight edges (*solid lines*). **A**: *Scenario 1*. The stimulus (*yellow arrow*) is compatible with a previously set attractor of the complex system encoded by a subset of the core nodes (*horizontal red double arrows*) and provokes a fast, matching response (*solid line yellow double arrows*), which transfers the system to this pre-set attractor. **B**: *Scenario 2*. The stimulus is incompatible with previously set attractors of core-nodes (*red*) provoking a fluctuation between attractors (*red double arrows*). Consequently, the stimulus is spread to the network periphery (*green nodes*), and induces a slow, system-level, integrative response (*dashed line yellow double arrows*). Here, a collective decision of the entire network emerges. **C**: *Scenario 3*. Repeated stimuli reconfigure the core (*red nodes*) encoding a new system attractor (*solid line yellow double arrows*). Overlapping subsets of core nodes may encode/connect multiple attractors.[14,86–90] The emergence of fast and slow responses is also illustrated by three pairs of videos of the additional supporting information on an illustrative network of neurons, as well as the real world social networks of network scientists[82] and high school students.[83]

# Additional supporting information

to

**The wisdom of networks: A general adaptation and learning mechanism of complex systems.** The network core triggers fast responses to known stimuli; innovations require the slow network periphery and are encoded by core-remodeling (http://onlinelibrary.wiley.com/wol1/doi/10.1002/bies.201700150/full)

*Peter Csermely (Department of Medical Chemistry, Semmelweis University, H-1094 Budapest, Hungary; Email: csermely.peter@med.semmelweis-univ.hu)*

This supporting information describes three video pairs that illustrate the (typically fast) retrieval and execution of previously encoded responses to known, "business as usual" situations that involve a few, highly connected nodes of the network core *versus* the (typically slow) development of novel responses to previously unknown, novel situations that involve a substantial number of network nodes from the network periphery. In the "business as usual" situation, core nodes have the same initial reaction (marked by identical colors), which soon becomes the general response of the entire network shifting the system to its respective attractor. In contrast, the initial response of core nodes in an unexpected situation may differ and may generate conflicts (as illustrated by the blinking different initial colors). In this "business as unusual" situation, the general response (illustrated by the final color) slowly emerges and involves the contribution of many individual, peripheral nodes, which represent the "wisdom of crowds".[1] In case of repeated stimuli this general response may be encoded as a novel attractor of the system by reconfiguring its core.

**1. A video pair that illustrates the execution of previously encoded, "fast" *versus* newly developing, "slow" responses in neuronal networks.** The video pair is an illustrative image-flow that indicates a putative activation series of neurons in the cases of "business as usual" ("fast") and unexpected ("slow") situations, respectively. The neuronal network of the videos was downloaded from the following site: http://topwalls.net/3d-graphics-network (retrieved on 08.19.2015).

- ➢ Movie S1. View the video of a "fast" decision-making process here: (http://onlinelibrary.wiley.com/wol1/doi/10.1002/bies.201700150/suppinfo) or here (http://linkgroup.hu/docs/video/FAST-decision-video-neurons.mp4)

- ➢ Movie S2. View the video of a "slow" decision-making process here: (http://onlinelibrary.wiley.com/wol1/doi/10.1002/bies.201700150/suppinfo) or here (http://linkgroup.hu/docs/video/SLOW-decision-video-neurons.mp4)

In the video pairs of the social networks of network scientists[2] and school children,[3] the vertical position of the network nodes marks their community centrality,[4] i.e., their importance within their network module. Nodes with the highest community centrality correspond to the "opinion leaders" of their community.[4] 2D network images were produced using the Moduland Cytoscape plug-in.[5] Video frames were made by creating a plug-in for the Blender software (Blender Foundation, Amsterdam, The Netherlands; https://www.blender.org/ retrieved on 08.22.2015) and were converted to a video using FFmpeg (an open-source multimedia system originated by Fabrice Bellard in 2000; https://www.ffmpeg.org/ retrieved on 08.22.2015).

**2. A video pair that illustrates the execution of previously encoded, "fast" *versus* newly developing, "slow" responses in a social network of network scientists.**[2] Here, the top nodes correspond to well-known members of the network science field.[4]

- ➢ Movie S3. View the video of a "fast" decision-making process here: (http://onlinelibrary.wiley.com/wol1/doi/10.1002/bies.201700150/suppinfo) or here (http://linkgroup.hu/docs/video/FAST-decision-video-scientists.mp4)

- ➢ Movie S4. View the video of a "slow" decision-making process here: (http://onlinelibrary.wiley.com/wol1/doi/10.1002/bies.201700150/suppinfo) or here (http://linkgroup.hu/docs/video/SLOW-decision-video-scientists.mp4)

**3. A video pair that illustrates the execution of previously encoded, "fast" *versus* newly developing, "slow" responses in a social network of school children.**[3] The social network is Community-44 of the Add Health survey, in which edge weights represent the strength of student friendships. This school community had four rather well-separated social communities of black and white, as well as lower and upper high school students.[3–5]

- ➢ Movie S5. View the video of a "fast" decision-making process here: (http://onlinelibrary.wiley.com/wol1/doi/10.1002/bies.201700150/suppinfo) or here (http://linkgroup.hu/docs/video/FAST-decision-video-students.mp4)

- ➢ Movie S6. View the video of a "slow" decision-making process here: (http://onlinelibrary.wiley.com/wol1/doi/10.1002/bies.201700150/suppinfo) or here (http://linkgroup.hu/docs/video/SLOW-decision-video-students.mp4)